\documentclass[aps,prl,superscriptaddress,twocolumn,showpacs]{revtex4}

\usepackage{amsmath, amssymb, amsthm, alltt, setspace,bbm}
\usepackage{graphicx}

\newcommand{\be}{\begin{equation}}
\newcommand{\ee}{\end{equation}}
\newcommand{\bs}{\begin{split}}
\newcommand{\es}{\end{split}}

\def\ket#1{\left\vert #1 \right\rangle}
\def\bra#1{\left\langle #1 \right\vert}

\DeclareMathOperator{\Tr}{\mathrm{Tr}}
\begin{document}

\title{A characterization of global entanglement}

\author{Peter J. Love}
\affiliation{D-Wave Systems Inc., 100-4401 Still Creek Drive, Burnaby, B.C.,   V5C 6G9 Canada}
\affiliation{Department of Mathematics, Tufts University, Bromfield--Pearson Building, Medford, MA 02155, USA}
\author{Alec \surname{Maassen van den Brink}}
\affiliation{D-Wave Systems Inc., 100-4401 Still Creek Drive, Burnaby, B.C.,   V5C 6G9 Canada}
\author{A. Yu. Smirnov}
\affiliation{D-Wave Systems Inc., 100-4401 Still Creek Drive, Burnaby, B.C.,   V5C 6G9 Canada}
\author{M.~H.~S.~Amin}
\affiliation{D-Wave Systems Inc., 100-4401 Still Creek Drive, Burnaby, B.C.,   V5C 6G9 Canada}
\author{M. Grajcar}
\affiliation{Institute for Physical High Technology, P.O. Box 100239, D-07702 Jena, Germany}
\affiliation{Frontier Research System, RIKEN, Wako-shi, Saitama, 351-0198, Japan}
\affiliation{Department of Solid State Physics, Comenius University, SK-84248, Bratislava, Slovakia}
\author{E. Il'ichev}
\affiliation{Institute for Physical High Technology, P.O. Box 100239, D-07702 Jena, Germany}
\author{A. Izmalkov}
\affiliation{Institute for Physical High Technology, P.O. Box 100239, D-07702 Jena, Germany}
\author{A. M. Zagoskin}
\affiliation{Frontier Research System, RIKEN, Wako-shi, Saitama, 351-0198, Japan}
\affiliation{Physics and Astronomy Dept., The University of British Columbia,6224 Agricultural Rd., Vancouver, B.C., V6T 1Z1 Canada}
\date{March 22, 2006}

\begin{abstract}
We define a set of $2^{n-1}-1$ entanglement monotones for $n$ qubits and give a single measure of entanglement in terms of these. This measure is zero except on globally entangled (fully inseparable) states. This measure is compared to the Meyer--Wallach measure for two, three, and four qubits. We determine the four-qubit state, symmetric under exchange of qubit labels, which maximizes this measure. It is also shown how the elementary monotones may be computed as a function of  observable quantities. We compute the magnitude of our measure for the ground state of the four-qubit superconducting experimental system investigated in [M.~Grajcar \emph{et al.}, Phys.\ Rev.\ Lett.\ \textbf{96}, 047006 (2006)], and thus confirm the presence of global entanglement in the ground state.
\end{abstract}

\maketitle

Entanglement is perhaps the fundamental property distinguishing quantum from classical physics. It underpins the violation of classical locally realistic descriptions of nature displayed by the Einstein--Podolsky--Rosen paradox, quantified by the Bell inequalities, and observed in the Aspect experiments~\cite{einstein1935,bell1966,aspect1981}. Entanglement also lies behind the greater information-theoretic power of quantum systems. Quantum computers require entanglement in order to exceed the capabilities of classical computers, as quantum computations without entanglement may be efficiently simulated classically~\cite{vidal2003}. Shared entangled states are also required for the implementation of quantum cryptography~\cite{gisin2001}.

In the context of quantum computation we wish to characterize the entanglement of states of $n$ two-level quantum systems (qubits). A (pure) state of $n$ qubits is called unentangled if its wave function may be written as an $n$-fold tensor product of states of the individual qubits. Any state which cannot be written in this form is entangled. A state is {\em globally entangled\/} (equivalently, {\em fully inseparable\/}) if it may not be written as a tensor product of states of any set of subsystems. This definition, however, does not give any insight into how strongly the qubits are entangled.

There are several approaches to the quantification of entanglement. Firstly, {\em canonical forms\/} for state vectors of qubit systems have been elucidated which represent locally inequivalent classes of states~\cite{bib:carteret,linden1999}. Knowledge of a set of canonical form parameters corresponds to complete knowledge of the entanglement properties of the state~\cite{linden1999}. However, the number of such parameters grows exponentially with the number of qubits, and their determination is nontrivial. Alternatively, {\em measures of entanglement\/} have been defined which are constant on locally equivalent states~\cite{Wootters1998,Meyer1998,Vidal2000,Meyer2002}. Such measures are the subject of the present paper. Finally, one may also define observables whose expectation values are positive on unentangled states and negative on entangled ones. Such {\em entanglement witnesses\/} have the advantage that they may be directly measured~\cite{bour2005}.

Measures of entanglement are real-valued functions on quantum states which are invariant under local unitary operations. They are also required to be {\em entanglement monotones}, that is, they must be non-increasing under local quantum operations combined with classical communication (LOCC)~\cite{Vidal2000}. For two-qubit entanglement there is only one nontrivial invariant, and the entanglement of two-qubit systems is therefore well understood for both pure and mixed states~\cite{Wootters1998}. The appropriate measure in this case is the concurrence, or functions thereof.

For more qubits the situation is substantially more complicated. The polynomial invariants of pure three-qubit states have been completely elucidated~\cite{Meyer1998} and the number of different types of such invariant identified for four qubits~\cite{Wallach2002}. For more than two qubits there is no single measure of entanglement, and therefore no unique maximally entangled state~\cite{dur2000}. The dimension of the space of invariants grows extremely rapidly with the number of qubits, indicating that the elucidation of a complete set of such invariants for $n$-qubit systems is not useful~\cite{Meyer1998}.

Detailed analysis of the three- and four-qubit cases led Meyer and Wallach to define a single scalar measure of pure-state entanglement~\cite{Meyer2002}.  This measure was further explored by Brennen~\cite{Brennen2003}, who showed that it is a monotone. The Meyer--Wallach (MW) measure written in the Brennen form is:
\be
  Q(\psi)=\frac{1}{n}\sum_{k=1}^n2\left(1-\Tr[\rho_k^2]\right),
\ee
where $\rho_k$ is the one-qubit reduced density matrix of the $k$th qubit after tracing out the rest.

The MW measure was originally described as a measure of global entanglement, to distinguish it from purely bipartite measures such as the concurrence. However, it is not able to distinguish states which are fully inseparable from states which, while entangled, are separable into states of some set of subsystems. This property becomes a serious drawback in the context of analysis of experimental data. For example, the MW measure is one both for the four-qubit GHZ state (a globally entangled four-qubit state~\cite{GHZ}) and a product of two two-qubit Bell states~\cite{Meyer2002}. This means that the measure is unable to distinguish an experiment preparing global four-qubit entanglement from one involving two qubits in a Bell state in Jena and two qubits in a Bell state in Vancouver.

The inability of the MW measure to distinguish sub-global and global entanglement arises for two reasons. Firstly, the summands measure the entanglement of each individual qubit with all the others. Secondly, the arithmetic average can obscure information about separability contained in the set of summands.

Scott generalized this measure for $n$ qubits divided into a set $S$ of $m$ qubits and a set $\bar S$ of $n-m$ qubits as follows:
\be
  Q_m(\psi) = \begin{pmatrix} n \\ m \end{pmatrix}^{\!\!-1}
  \sum_{|S|=m}\frac{2^m}{2^m-1}\left(1-\Tr[\rho_S^2]\right),
\ee
where $\rho_S=\Tr_{\bar S}[\rho]$ is the $m$-qubit reduced density matrix of the qubits in subset $S$ after tracing out the rest, and $|S|$ is the number of qubits in~$S$. There are $\lfloor n/2\rfloor$ distinct measures because $\Tr[\rho_S^2]=\Tr[\rho_{\bar S}^2]$ (as may be shown using, e.g., a Schmidt decomposition~\cite{bib:niechuang}), and so $(1{-}2^{-m})Q_m=(1{-}2^{m-n})Q_{n-m}$. The Scott measure subsumes the MW measure as the special case $m=1$~\cite{Scott2004}.

In this Letter we propose a characterization of entanglement based on a set of monotones, each of which corresponds to a bipartition of the qubits. We define a scalar measure of {\em global\/} entanglement which is nonzero only on fully inseparable states. We show that this measure is maximized by a highly entangled four-qubit state, and compute this measure for the ground state of an experimentally determined Hamiltonian.

The above measures of entanglement all take the form of arithmetic averages of quantities
\be\label{elements}
  \eta_S=\frac{2^{|S|}}{2^{|S|}-1}\left(1-\Tr[\rho_S^2]\right).
\ee
Each $\eta_S$ ($0\leq\eta_S\leq1$) characterizes the entanglement of the qubits in $S$ with the rest. For a pure state $\rho$, the reduced density matrices $\rho_S$ and $\rho_{\bar S}$ are pure if and only if $\rho$ is separable with respect to the division $(S,\bar S)$. As a consequence each $\eta_S$ is zero if and only if the state $\rho$ is separable with respect to the partition of the qubits into $S$ and $\bar S$.

Vidal showed that any unitarily invariant concave function of a partial trace is an entanglement monotone~\cite{Vidal2000}. The function $1-\Tr[x^2]$ is concave. Denoting the unitary group acting on the Hilbert space of the set of qubits $S$ by~$U_S$, we see that each $\eta_S$ is invariant under any unitary in $U_S\otimes U_{\bar S}$. The intersection of all groups $U_S\otimes U_{\bar S}$ for all $S$ is the local unitaries, $U(2)^{\otimes n}$. Hence the entire set of~(\ref{elements}) is invariant under local unitaries and each element of the set is a monotone.

We propose the set of quantities
\be\label{eq:mset}
  \{\eta_S|\:|S|=m,\: 1\leq m \leq \lfloor n/2\rfloor\}
\ee
as a useful characterization of entanglement. Since $(1{-}2^{-|S|})\eta_S=(1{-}2^{-|\bar{S}|})\eta_{\bar{S}}$, there are $2^{n-1}-1$ distinct measures and so computation of the entire set is not convenient for large numbers of qubits. However, this characterization will remain practical for experimental systems in the near future. We can also define functions of this set which satisfy additional desiderata beyond those required for entanglement monotones. The MW measure and Scott's generalizations are then special cases where the relevant functions of this set are taken to be arithmetic averages of specific subsets.

There is a special case in which the characterization of entanglement given by~(\ref{eq:mset}) is compact enough for practical purposes for an arbitrary number of qubits. For pure quantum states which only change by a global phase under exchange of qubit labels, all $\eta_S$ with $S$ containing the same number of qubits are equal. In this case there are simply $\lfloor n/2\rfloor$ distinct measures $\eta_S$---a manageable characterization of entanglement. This special case includes bosonic, fermionic and anyonic state vectors of interest in many-body physics. In general, the states occurring in quantum information theory do not possess such exchange symmetry. Namely, qubits are typically grouped into registers fulfilling different functions, and hence must be distinguishable. However, it is notable that the entangled $n$-qubit W and GHZ states are exchange invariant~\cite{dur2000,GHZ}.

As an example of a function of the $\eta_S$ which is nonzero only on globally entangled states we define:
\be\label{measure}
  {\cal R}(\psi)=\Biggl({\prod_{1\le|S|\le|\bar{S}|}}^{\mskip-23mu\prime} \;\eta_S(\psi)\Biggr)^{\!\!1/(2^{n{-}1}-1)},
\ee
where the prime indicates that if $|S|=n/2$, each partition is also included only once. This is simply the geometric average of all distinct $\eta_S$. This is non-increasing under any set of operations for which the $\eta_S$ are, and vanishes on any state which is not globally entangled; see below (\ref{eq:conroof}) for details.

We now compare $\cal R$ to the MW measure. For two qubits only a single bipartition exists, and ${\cal R}$ reduces to the Meyer--Wallach measure. The concurrence is simply~$\sqrt{{\cal R}}$. For three qubits there are three bipartitions and two different types of globally entangled states, the GHZ type and the W~type~\cite{dur2000}. The measures $\eta_1$, $\eta_2$, $\eta_3$, and ${\cal R}$ are all equal to $1$ for the GHZ state and $\frac{8}{9}$ for the W state, the same values as the MW measure.

The case of four qubits is qualitatively different. Four is the smallest number of qubits for which there are two distinct types of partition, namely one-plus-three and two-plus-two. While the four-qubit GHZ state maximizes $\eta_{1}(=\eta_2=\eta_3=\eta_4)=1$, the value of the two-plus-two measure is only~$\frac{2}{3}$. The four-qubit W state has $\eta_{1}(=\eta_2=\eta_3=\eta_4)=\frac{3}{4}$, and identical two-plus-two entanglement to the GHZ state. Are there exchange-symmetric states for which $\eta_{1{+}3}=1$ and $\eta_{2+2}>\frac{2}{3}$? Given that a choice of a single scalar entanglement measure is also a selection of the states which maximize this function, maximization of ${\cal R}$ determines whether there exist states with greater entanglement than the GHZ state by our definition.

The determination of the maximum of ${\cal R}$ for four qubits is nontrivial. The search space of pure four-qubit states has thirty real dimensions---much too large for a naive maximization of the complicated objective function~${\cal R}$. Because we are only interested in searching a subspace of locally inequivalent states, we can utilize a canonical form with only nineteen real parameters~\cite{bib:carteret}. This is still too large for brute-force search, but by further restricting our search with the Ansatz that the states of interest be exchange symmetric we may construct a six-parameter normal form
\be
  \ket{\psi}=l_0e^{i\phi_0}\ket{0000} + 2l_1\ket{\overline{W}_4} + \sqrt{6}l_2e^{i\phi_1}\ket{V_4} + l_3\ket{1111},
\ee
where $\ket{\overline{W}_4}$ is the four-qubit W state with all bits flipped and $\ket{V_4}$ is the uniform superposition over all logical basis states with exactly two qubits in state one. All parameters are real, and $l_1$, $l_3$ are non-negative.

A numerical search over the six parameters identified
\be
  \ket{\psi_\mathrm{m}}=\frac{1}{\sqrt{3}}(\ket{0000} + \sqrt{2}\ket{\overline{W}_4})
\ee
as an exchange-symmetric state which maximizes~${\cal R}$. The state  $\ket{\psi_\mathrm{m}}$ (and its local equivalents) has $\eta_1=\eta_2=\eta_3=\eta_4=1$ and $\eta_{12}=\eta_{13}=\eta_{14}=\frac{8}{9}$, giving ${\cal R}\simeq0.95077$. Hence the one-plus-three entanglement of this state is identical to that of the GHZ state, but the two-plus-two entanglement is larger than that of either the GHZ or W state. Exhibition of $\ket{\psi_\mathrm{m}}$ illustrates the utility of our entanglement characterization, as the MW measure is unable to distinguish states with the same values of~$\eta_1$. This state was also identified by analytical methods as a maximally entangled four-qubit state in~\cite{osterloh2005}.

\begin{figure}[t]
\centering
\includegraphics[width=7cm]{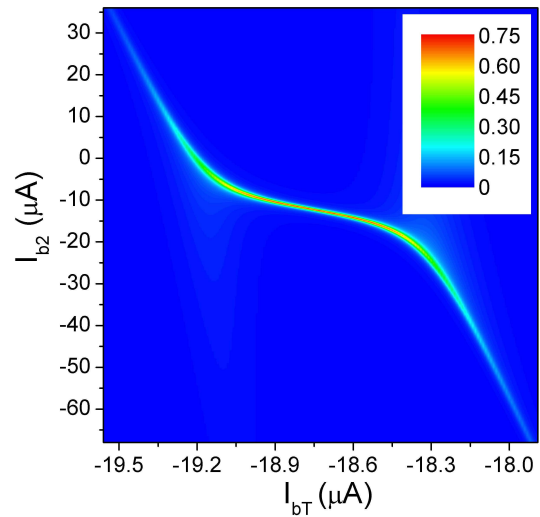}
\caption{Entanglement ${\cal R}(\psi)$ of the ground state of the four-qubit experimental system described in~\cite{grajcar2005}. The abcissa is the global bias applied to all qubits, and the ordinate is the bias current applied to qubit two. The maximum entanglement value is $0.75\pm 0.05$.}\label{fig1}
\end{figure}

The definition of ${\cal R}$ was motivated by the desire to characterize the entanglement present in the ground state of the system of four coupled superconducting three-junction flux qubits described in~\cite{grajcar2005}. The low-energy effective Hamiltonian of the qubit system was determined by impedance measurement~\cite{greenberg2002}. The ground state of this Hamiltonian was then found numerically, and its entanglement computed; see Fig.~\ref{fig1}. The maximum value is $0.75\pm 0.05$ at a point where the ground state is close to $\frac{1}{\sqrt{2}}\{\ket{0101}+\ket{1010}\}$ (locally equivalent to the four-qubit GHZ state), for which ${\cal R}=0.8405$. This confirms the existence of substantial global entanglement in the ground state, a prerequisite for adiabatic quantum computation~\cite{Farhi}.

%trim trims off from l b r t
%,width=0.48\textwidth,height=0.21\textheight
%more

The experiment in~\cite{grajcar2005} determined the ground state, and our analysis computed ${\cal R}$ {\em ex post facto}. Alternatively, one could use quantum state tomography~\cite{bib:niechuang} to determine the state vector and again compute ${\cal R}$ from this state. Can the set of quantities~(\ref{eq:mset}) be determined from experimentally observable quantities directly?

We address this question by writing our density matrix in the extended Pauli basis of $n$-fold tensor products of the Pauli matrices and the identity. We label each element of this basis with a word $L$ of length $n$ on the alphabet $A=\{0,1,2,3\}$. Here, zero labels the $2\times2$ identity and $1$, $2$, $3$ label the Pauli matrices. Denote the set of labels by $A^n=\{0,1,2,3\}^n$, and let $L_i$ be the $i$th letter of $L\in A^n$. The basis element corresponding to $L$ is $\sigma_L=\sigma_{L_n}\otimes\dots\otimes\sigma_{L_1}$. By slight abuse of notation, we write the word of all zeroes (which labels the $2^n\times2^n$ identity matrix) as $0^n$.

Any $n$-qubit density matrix may be written in this basis:
$\rho=2^{-n}\sum_{L\in A^n}\alpha_L\sigma_L$,
where $\alpha_{0^n}=1$ and each $\alpha_L$ is the expectation value of the observable~$\sigma_L$. Consider the reduced density matrix arising from tracing out the $k$th qubit; only terms for which $\sigma_{L_k}=\sigma_0$ will appear. One quarter of the words have $L_k=0$, so we obtain:
\be
  \Tr_k[\rho]=\frac{1}{2^{n-1}}
  \sum_L\delta_{L_k,0}\,\alpha_{L}\sigma_{L\setminus \{k\}},
\ee
where $L\setminus \{k\}$ is the word obtained from $L$ by deleting the $k$th letter and $\delta_{i,j}$ is the Kronecker delta. Similarly, the reduced density matrix $\rho_S$ is given by:
\be
  \rho_S=\frac{1}{2^{|S|}}\biggl(\sigma_{0^{|S|}}+\sum_{L\in B_S\subset A^n}\alpha_L\sigma_{L\setminus\bar{S}}\biggr),
\ee
where the set of Pauli basis coefficients appearing in the sum is $B_S=\{L \in A^n  | L_k=0, \forall k\in{\bar S}\bigwedge L\neq 0^{|S|}\}$. The measures of entanglement $\eta_S$ have a particularly simple form in terms of the $\alpha_L$:
\be\label{eq:obsent}
  \eta_S=1-\frac{1}{2^{|S|}-1}\sum_{L\in B_S}\alpha_L^2.
\ee
This simple form confirms the suitability of the extended Pauli basis for consideration of entanglement~\cite{bib:realdenmat}. The above also clarifies the connection between the approach taken here and the invariants described in~\cite{aschauer2003}.

The above shows that the $\eta_S$ can indeed be determined from observable quantities. For example, for four qubits, determination of $\eta_{12}$ requires measuring the 15 observables  $\sigma_a\otimes\sigma_b\otimes\sigma_0\otimes\sigma_0$, $(a,b)\neq(0,0)$. The expression~(\ref{eq:obsent}) for the $\eta_S$ also implies that a smaller set of measurements can bound the entanglement. We also note that any special properties of the state being measured (such as exchange symmetry) may be exploited to further reduce the number of measurements required.

The discussion above considered only pure states. However, any monotone $Q(\psi)$ may be extended to a monotone on mixed states $\tilde{Q}(\rho)$ by the convex-roof construction~\cite{Vidal2000}:
\be\label{eq:conroof}
  \tilde{Q}(\rho) = \min_{\Upsilon} \sum_i p_i\, Q(\psi_i).
\ee
$\Upsilon$ is the set of ensembles realizing the density matrix~$\rho$: $\Upsilon = \{\{p_i, \ket{\psi_i}\}| \sum_ip_i\ket{\psi_i}\bra{\psi_i} =\rho\}$. First, we apply (\ref{eq:conroof}) to the $\eta_S$ [which were argued above (\ref{eq:mset}) to be well-defined pure-state monotones]. One verifies that $\tilde{\eta}_S(\rho)=0$ if and only if $\rho$ is separable under $(S,\bar{S})$. [Naive evaluation of (\ref{elements}) might yield a nonzero value even for separable~$\rho$.] Subsequently, $\mathcal{R}$ is defined by (\ref{measure}) with $\eta_S\mapsto\tilde{\eta}_S$ even in the mixed case. Since all of its factors are non-increasing under LOCC, the same is true for~$\mathcal{R}$. Moreover, $\mathcal{R}(\rho)=0$ if and only if one of its factors vanishes, which is equivalent to separability by the above.

The characterization of global entanglement proposed here is based on consideration of all bipartite separations. It cannot be described as complete, as the number of measures is less than the number of parameters of the appropriate canonical form. This poses the question: what distinguishes two locally inequivalent states with equal values of all $\eta_S$? Answering this question could motivate new measures of entanglement based on concepts beyond separability, as has already been considered in~\cite{paz-silva2006}.

\section*{Acknowledgements} AZ acknowledges support by the NSERC Discovery Grants Program. PJL and AMvdB would like to thank J.~D. Biamonte, Andreas Osterloh, and Jens Siewert for helpful discussions, and Otfried G\"uhne for pointing out the relevance of~\cite{aschauer2003}.

\end{document}